\pdfoutput=1

\documentclass[11pt]{article}

\usepackage[final]{acl}

\usepackage{times}
\usepackage{latexsym}

\usepackage[T1]{fontenc}

\usepackage[utf8]{inputenc}

\usepackage{microtype}

\usepackage{inconsolata}

\usepackage{graphicx}

\usepackage{enumitem}
\usepackage{mathtools}
\usepackage{multirow}
\usepackage{booktabs}
\usepackage{svg}

\title{TopoLedgerBERT: Topological Learning of Ledger Description Embeddings using Siamese BERT-Networks.}

\author{
 \textbf{Sander Noels\textsuperscript{1,2}},
 \textbf{Sébastien Viaene\textsuperscript{1}},
 \textbf{Tijl De Bie\textsuperscript{2}}
\\
\\
 \textsuperscript{1}Silverfin, Gaston Crommenlaan 12, 9050 Ghent, Belgium \\
 \textsuperscript{2}Department of Electronics and Information Systems, Ghent University, 9000 Ghent, Belgium
\\
 \small{
   \textbf{Correspondence:} \href{mailto:sander.noels@ugent.be}{sander.noels@ugent.be}
 }
}

\begin{document}
\maketitle
\begin{abstract}
This paper addresses a long-standing problem in the field of accounting: mapping company-specific ledger accounts to a standardized chart of accounts. We propose a novel solution, TopoLedgerBERT, a unique sentence embedding method devised specifically for ledger account mapping. This model integrates hierarchical information from the charts of accounts into the sentence embedding process, aiming to accurately capture both the semantic similarity and the hierarchical structure of the ledger accounts. In addition, we introduce a data augmentation strategy that enriches the training data and, as a result, increases the performance of our proposed model. Compared to benchmark methods, TopoLedgerBERT demonstrates superior performance in terms of accuracy and mean reciprocal rank.
\end{abstract}

\section{Introduction}
\label{sec:introduction}

Ledger accounts are essentially the building blocks of accounting that provide detailed records of all financial activities within a company. They collectively form a hierarchical structure known as the chart of accounts (COA), which offers a comprehensive view of a company's financial activities. However, companies often customize their COA to accommodate their specific business needs and regulatory requirements, resulting in a lack of standardization and comparability \citep{jorgensen2021machine,wang2023standardizing}. This scenario has brought about various standardization initiatives aiming to enhance the quality of COAs. A standardized COA not only improves the accuracy of financial data, but also promotes comparability and improves the overall quality of the financial information reported \cite{rylee2017so}. Moreover, it paves the way for machine learning solutions that can generalize financial information between companies \cite{bergdorf2018machine}.

The mapping of company-specific ledger accounts to a common COA remains a challenge. Existing academic and industry solutions typically create custom machine learning models for individual companies, or limit data to a predefined set of accounts \cite{bergdorf2018machine}. Furthermore, the hierarchical information inherent in a COA, which could significantly enhance the performance of the mapping solution, has never been used.

We hypothesize that the hierarchical structure of the ledger accounts within the COA and their interrelationships can enhance the performance of the mapping solution. Considering the meaning and hierarchical nature of the ledger account descriptions, we believe that a model that understands these relationships should deliver superior ledger mapping results.

We propose TopoLedgerBERT, a novel sentence embedding method for ledger account mapping that overcomes the limitations of existing methods by exploiting the hierarchical nature of ledger accounts. To create the mapper, we take a ledger description from the custom COA, compute its embedding, and then calculate its distance to embeddings of standard ledger accounts. The standard ledger account having the minimum distance is considered the mapping for the custom ledger account. TopoLedgerBERT incorporates a graph distance concept to include hierarchical information from the COA during the sentence embedding process. By doing so, it captures not only the semantic similarity of ledger accounts but also their hierarchical structure. Additionally, this method is capable of making predictions for previously unseen companies and COAs.

\noindent Our main contributions are summarized as follows:
\begin{itemize}[nosep]
\item We propose TopoLedgerBERT, a novel sentence embedding method specifically designed for the mapping of ledger accounts in accounting. 
\item Unlike existing solutions that require customized machine learning models for individual companies, TopoLedgerBERT can adapt to new companies and COAs without any prior setup.
\item We demonstrate that the TopoLedgerBERT model outperforms benchmark methods in terms of accuracy and mean reciprocal rank.
\item Our empirical evaluation reveals that TopoLedgerBERT effectively captures hierarchical relationships between different ledger accounts.
\item Finally, we demonstrate the importance of fine-tuning pre-trained embeddings and employing data augmentation strategies for improving the model's performance.
\end{itemize}

The remainder of this paper is as follows: Section \ref{sec:related-work} reviews the related work. Section \ref{sec:background} delves into the background and explains how a COA can be represented as a graph. Section \ref{sec:topoledgerbert} details our proposed TopoLedgerBERT model, followed by an empirical evaluation in Section \ref{sec:evaluation}. Finally, Section \ref{sec:conclusion} concludes the paper.

\section{Related Work}
\label{sec:related-work}

\subsection{Ledger Accounts and the Chart of Accounts}

Ledger accounts serve as units that record and summarize a company's transactions. A ledger account typically includes an ID that possibly reflects the hierarchy and a short description that can sometimes be challenging to decipher. Each ledger account also contains a set of bookings for each financial year. A company generally comprises around 100 ledger accounts.

These accounts collectively form a hierarchy known as the "chart of accounts" (COA), a classification system for financial information. The COA segregates expenditures, revenues, equity, assets, and liabilities into distinct categories, providing a comprehensive breakdown of all financial transactions during a specific accounting period. However, a long-standing issue in this domain is the lack of standardization in the COA. 

\subsection{Factors Driving the Customization of COAs}

Customizing COAs is driven by several factors. Companies often require specialized ledger accounts to support specific financial transactions relevant to their sector \cite{jorge2022new}. In addition to sector-specific needs, the complexities of tax law and individual company requirements contribute to the customization of COAs. These complexities underline the subjective nature of bookkeeping, where companies vary in their practices, making standardization a challenging endeavor \cite{liu2021categorization}. Two examples can illustrate these complexities:

\begin{itemize}[nosep]
\item \textbf{Tax Compliance}: US tax laws dictate varying rates at which different expenses are deductible. For example, expenses for business vehicles and office snacks are 100\% deductible, while business meals are only 50\% deductible. Consequently, separate accounts for these expenses become necessary for accurate tracking and compliance.
\item \textbf{Business Needs}: Companies may have different needs for account specificity. For example, company A might be satisfied with a generalized account named "cars and trucks" for all its automobile-related expenses. However, company B might prefer more granular accounting and create separate accounts for "fuel", "vehicle maintenance", and "vehicle insurance".
\end{itemize}

Although customization can support specific business needs and regulatory compliance, it can also lead to inconsistencies and difficulties in comparing financial statements between companies \cite{dhole2015effects,noels2022earth,noels2023efficient}. Therefore, while supporting the freedom for companies to customize their COAs, it is essential to develop methods for translating these custom COAs into a unified structure. 

\subsection{The Importance of Standardization}

Standardization initiatives have been introduced, both on national and international scale, to improve various aspects of accounting \cite{eurostat2017,jorge2022new}. For example, in 2008, the European Commission recommended the use of standard COAs for small enterprises in a report that highlights best accounting practices \cite{EuropeanCommission2008}. Standardization initiatives are beneficial for various reasons. Primarily, they enhance the accuracy of financial data and, consequently, financial reporting. A unified COA allows for the uniform sorting and aggregating of financial data, facilitating consistent accounting practices and routines. This, in turn, helps to develop effective IT and control systems \cite{jorge2022new}. According to \citet{rylee2017so}, the standardization of COAs ensures a consistent and streamlined reporting of financial data, thus preventing inaccuracies that might result from too complex COAs.

\subsection{Standardization Initiatives in Related Work}

The "multiple charts of accounts problem" is a long-standing issue in both academia and industry due to the unique customization of COAs by different companies. The challenge lies in the need for a system that can effectively generalize and merge financial information across various companies and sectors, even with minimal historical data \cite{bergdorf2018machine}.

Past strategies have tried to circumvent this problem using company-specific classifiers or the use of uniform ledger accounts or COAs. \citet{jorgensen2021machine} propose a system capable of mapping transactions to the appropriate ledger accounts across different companies. However, this approach required one to first manually map specific account codes into a unified space. \citet{munoz2022hierarchical} limit themselves to a set of accounts used by different companies, employing hierarchical classifiers to predict the correct account code for specified invoice line items. Despite the innovative aspect of their approach, this method is limited to the account codes present in their training data, making it ineffective for accounts added after their training period.
\citet{noels2022earth,noels2023efficient} define a COA with all possible financial accounts hierarchically structured within a financial statement, allowing them to represent every company. This allows them to compute the similarity between the financial statements of each company. However, due to the lack of uniformity, the use of this method for automated financial statement analysis remains limited, indicating the need for a method capable of automating the translation to a standardized COA.

All the methods discussed face the challenge known as the "cold-start problem", which occurs when a system has to make predictions for ledger accounts that were not present in the training dataset. This limitation has inspired efforts toward the automation of COA standardization. In response to this challenge, a viable solution lies in automating the translation of a company's unique financial structure into a unified accounting chart. Such a chart should be capable of accounting for all potential company transactions and could therefore be effectively implemented in machine learning systems. According to \citet{jorgensen2021machine}, the most significant opportunity for improvement lies in the automated transformation to a standardized COA, while still allowing company-specific COAs tailored to individual company needs or tax regulations.

\subsection{Account Mapping in Academia and Industry} 

Companies like Oracle and Silverfin provide their own mapping solutions. Oracle's solution includes a COA mapping feature for data consolidation, but this rule-based system requires substantial maintenance due to the need for continuous adaptation. Silverfin uses machine learning in its AI mapping feature for better adaptability, though it struggles with new COAs. Both aim to automate the process of mapping custom ledger accounts, highlighting their importance in the industry.

There is a growing academic interest in automated account mapping solutions, as demonstrated by a system proposed by \citet{wang2023standardizing}. The system standardizes custom ledger accounts by automatically mapping them to standard ones using three pre-trained embedding methods (TF-IDF, Word2Vec, FinBERT). To the best of our knowledge, this is the first academic work that offers a solution to the laborious process of mapping custom ledger accounts to a standardized taxonomy.

The rise of large language models opens up new opportunities in financial data applications. The rationale is that these methods could be better adapted to the unique terminology used in financial data, potentially improving the accuracy of mapping custom ledger accounts \cite{noels2023automated}. According to \citet{wu2023bloomberggpt}, the complexity of the financial domain, the unique terminology, scarce data, and privacy regulations have limited the success of large language models in providing valuable insights. Therefore, fine-tuning pre-trained embeddings, such as FinBERT, could potentially overcome these hurdles and offer more accurate representations of financial data, thus improving the mapping process \cite{liu2021finbert}.

\citet{liu2021categorization} introduce another promising approach, using siamese network-inspired neural networks to perform binary matching of transaction and account information. This method uses pre-trained embedding models to place transactions and accounts in the same vector space. This approach could be applied to the account mapping problem, where custom ledger account descriptions are paired with their standardized ledger account descriptions. This approach offers a distinct advantage over classification-based methods, as it moves away from handling a fixed COA. 

In conclusion, the siamese network-inspired approach appears to be a promising avenue for future research, as it is not restricted to a predefined number of accounts and can accommodate unseen ledger accounts and COAs. Additionally, the idea of fine-tuning language models on financial data presents a compelling opportunity. Given the unique terminology, models trained specifically for this mapping problem would be better equipped to handle the complexity and nuances of ledger descriptions. These insights inspire further exploration of the development of more robust, adaptable, and accurate systems to automate the account mapping process in financial accounting.

\section{Background}
\label{sec:background}

This section provides an overview of the COA as a graph and the application of siamese BERT-networks in account mapping.

\subsection{Chart of Accounts as a Graph}
\label{subsec:coa-as-graph}

As stated by \citet{yang2013balance}, a vertex-labeled tree is a natural representation of the ledger accounts present within a COA. 

\begin{figure}[htbp]
  \includegraphics[width=\columnwidth]{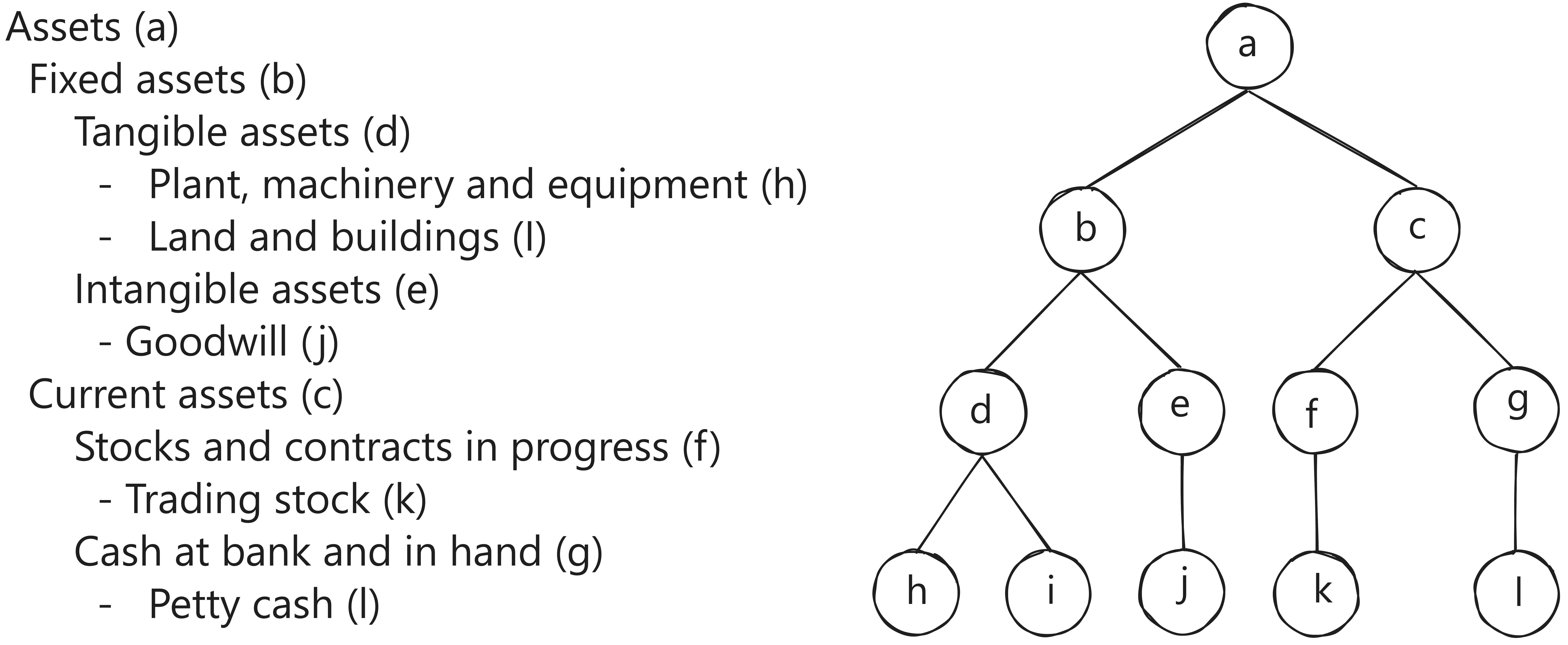}
  \caption{Left: Assets subsection of the balance sheet. Right: A vertex-labeled tree representation of the assets subsection of the balance sheet.}
  \label{fig1}
\end{figure}

As an example, we consider the \emph{assets} section of a balance sheet. The \emph{assets} section can be divided into \emph{fixed} and \emph{current assets}. Both of these ledger accounts can be subdivided into more detailed accounts, as shown in Figure \ref{fig1}. Note that the vertex-labeled representation of a COA is not limited to this specific example. A subset of ledger accounts and their reciprocal relationship are given for exemplary purposes. 

This representation method clearly preserves the structural property of a COA. This means that a COA for every company can be represented by a vertex-labeled tree, where the vertex labels are the ledger account descriptions. 

\subsection{Siamese BERT-Networks}

Siamese BERT-networks are a type of sentence embedding method that can embed a sentence, taking into account the semantic and contextual information of a sentence. In the context of account mapping, these networks use the bi-encoder transformer architecture of the Sentence-BERT network \cite{reimers2019sentence} to learn a joint representation space of the ledger account descriptions and their standardized counterparts.

Training a sentence embedding method involves inputting pairs of sentences into the network. These pairs can be either positive pairs, which consist of semantically similar sentences, or negative pairs, which consist of semantically dissimilar sentences. The network then calculates the distance between the two inputs. It attempts to minimize the distance for similar inputs (e.g., identical account descriptions) and maximize the distance for dissimilar inputs (e.g., different account descriptions). 

\section{TopoLedgerBERT}
\label{sec:topoledgerbert}

\subsection{Rationale Behind the Proposed Solution}

The rationale behind the proposed solution is to address the limitations of current sentence embedding methods in the context of ledger account mapping. As demonstrated in previous studies, the hierarchical relationships between the ledger accounts play a significant role in understanding the financial position of a company \cite{yang2013balance,noels2022earth,noels2023efficient}. Therefore, the embedding method we propose aims to capture not only the semantic similarity of the ledger accounts, but also their hierarchical composition within the COA. In doing so, we hypothesize that the proposed method can provide better embeddings for ledger accounts, leading to improved mapping accuracy and efficiency in the accounting domain.

Let us consider a specific example. The ledger account for \textit{land and buildings} can be subdivided into \textit{agricultural land} and \textit{residential land}. In this context, the proposed method should consider \textit{agricultural land} and \textit{residential land} as more similar to each other than to other ledger accounts that are located farther apart on the graph. This is due to their direct hierarchical relationship and position in the COA. 

\subsection{TopoLedgerBERT}

As mentioned in Section \ref{subsec:coa-as-graph}, a COA can be symbolized by a vertex-labeled tree, represented as $T = (V,E,L, \phi)$. In this context, $V(T) = \{1, \ldots, n\}$ stands for a set of $n$ vertices, $E \subseteq V \times V$ indicates the set of edges, and $L$ is a collection of unique labels corresponding to each vertex in $V$. These labels in $L$ define the standardized ledger account descriptions within a target COA. Therefore, we introduce a function $\phi: V\to L$ that maps every vertex $v$ in $V$ to a unique label $l$ in $L$.

We also define the shortest path distance $d(i, j)$ between the vertices $i$ and $j$ in $T$ and construct the distance matrix $D(T)$ of $T$. Using $D(T)$, we construct a similarity matrix $S(T)$, where each value 
\begin{equation}
s_{ij} = 1 - \frac{d_{ij}}{\max(D)}
\end{equation}
represents the similarity between the vertices $i$ and $j$. If the value of $s_{ij}$ is high, it signifies close proximity within the tree $T$, while low values indicate a significant distance.

\begin{figure}[htbp]
    \centering
    \includegraphics[width=0.7\columnwidth]{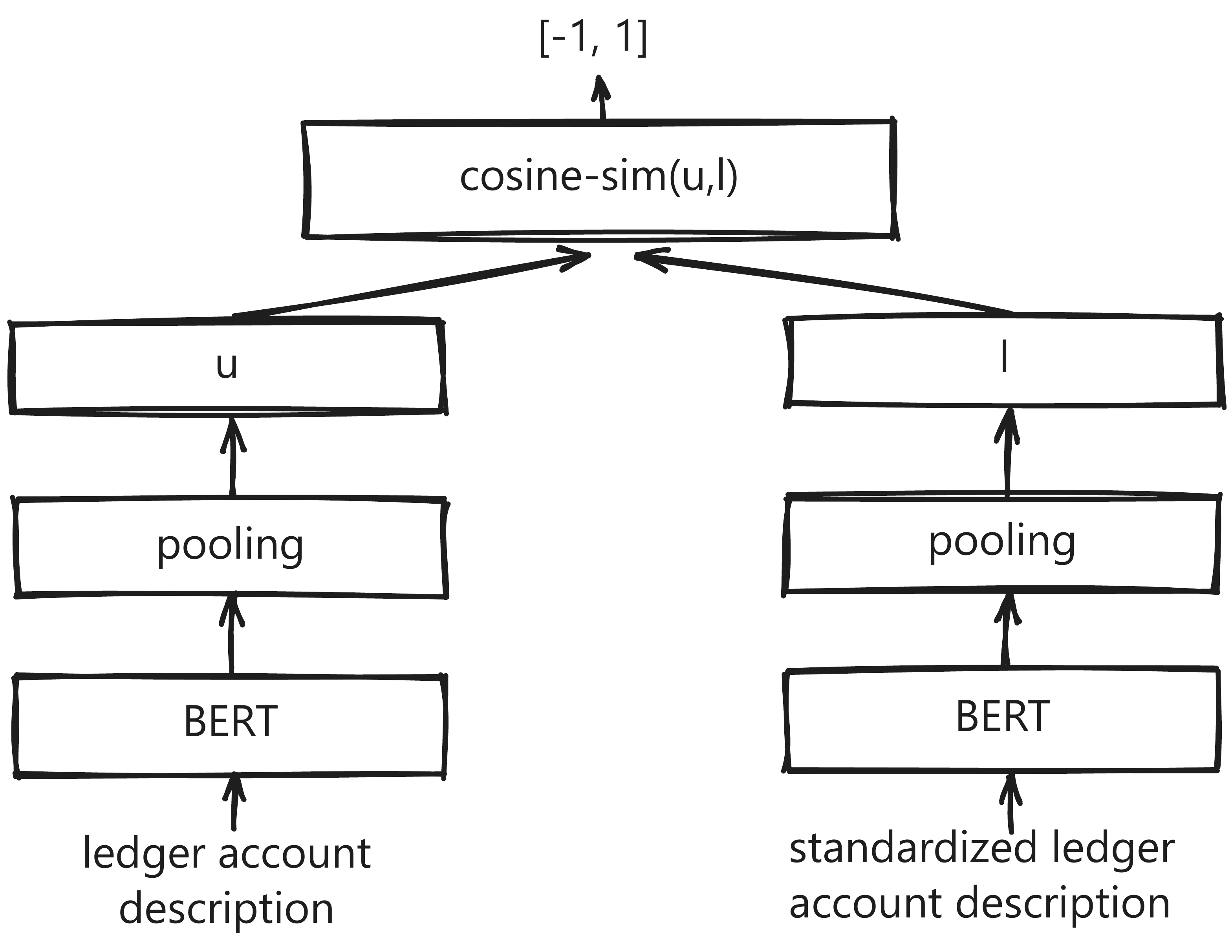}
    \caption{Diagram of the Sentence-BERT architecture for computing ledger account description similarity scores for the ledger account mapping problem.}
    \label{fig2}
\end{figure}

We create positive training samples by pairing each original ledger account description with its corresponding ledger account from the COA, assuming that there is an appropriate standardized ledger account for each account description. We denote the set of original ledger account descriptions as $U$, and the corresponding standardized ledger account descriptions as $L$. Each sample in the positive training dataset $D_{+}$ is a triplet $(u, l_u, s_i) \in U \times L \times \{1\}$, where $s_i = 1$ indicates the highest similarity score. 

To define the set $D_{+}$, we introduce $v_u$ which represents the vertex in $V$ corresponding to the original ledger account description $u$. The function $\phi: V \rightarrow L$ then maps each vertex $v_u$ to its corresponding unique label $l_u$. Given these definitions, we can formally define $D_{+}$ as follows:
\begin{equation}
D_{+} = \{(u, \phi(v_u), 1) | u \in U\},
\end{equation}
where $u$ represents an original ledger account description, and $\phi(v_u)$ denotes its corresponding label in the COA.

For each tuple $(u, \phi(v_u), 1) \in D_{+}$, we generate $K$ additional negative training samples. First, let us define the set $V^K_{\setminus v_u} \subseteq V\setminus\{v_u\}$ with $|V^K_{\setminus v_u}| = K$, a randomly selected subset of $K$ vertices from $V$ different from $v_u$. Subsequently, for each $v \in V^K_{\setminus v_u}$, we calculate the similarity score $s_{v_{u}v}$ using the similarity matrix $S(T)$. The set of negative training samples $D_{-}$ is defined as: 
\begin{equation}
D_{-} = \{(u, \phi(v), s_{v_{u}v}) | u \in U, v \in V^K_{\setminus v_u}\}, 
\end{equation}
with $|D_{-}| = K|D_{+}|$.

\begin{figure*}[htbp]
    \includegraphics[width=\textwidth]{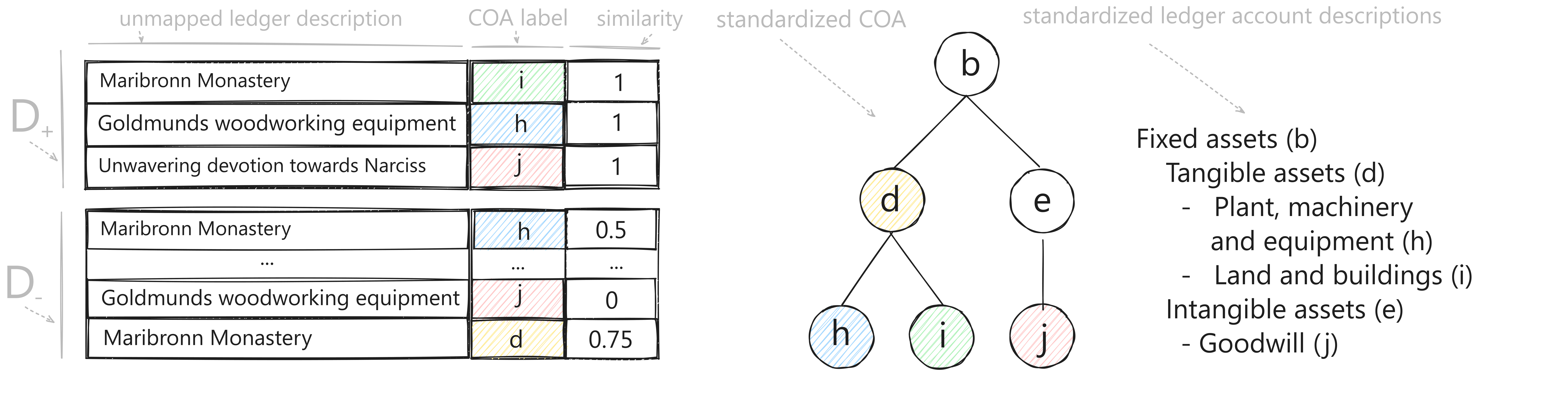}
    \caption{Example construction of an augmented dataset, $D_{aug}$, by TopoLedgerBERT for ledger account mapping.}
    \label{fig3}
\end{figure*}

Finally, the augmented data set $D_{aug}$ is the union of $D_{+}$ and $D_{-}$:
\begin{equation}
D_{aug} = D_{+} \cup D_{-}.
\end{equation}
$D_{aug}$ functions as the input for the TopoLedgerBERT model, as shown in Figure \ref{fig2}. The dataset augmentation method is visualized in Figure \ref{fig3}. 

\section{Empirical Evaluation}
\label{sec:evaluation}

This section provides a thorough empirical evaluation of TopoLedgerBERT's effectiveness in mapping ledger account descriptions to a standardized taxonomy, based on experiments using real-world financial data.

\subsection{Data Description}

The TopoLedgerBERT model uses a confidential dataset from Silverfin\footnote{\url{https://www.silverfin.com}}, an accountancy cloud service provider. This dataset comprises financial statement data from 5,608 UK companies from various industrial sectors and market sizes. Each company has mapped their original ledger account descriptions to the corresponding ledger accounts of one of six possible COAs, resulting in 521,265 unique mappings. The dataset is split into a 90\%-10\% training and testing set. 

For each unique COA in the dataset, a vertex-labeled tree $T_{config} = (V_{config},E_{config},L_{config},\phi_{config})$ with $config \in \{1,2,3,4,5,6\}$ was constructed. Each chart has $\{282467,151524,40183,28573,10311,8207\}$ mappings, respectively. 

\subsection{Methods Evaluated}

To assess the performance of our proposed TopoLedgerBERT model, we benchmark it against several established models. The following methods are considered:

\begin{itemize}[nosep]

\item Standard SBERT: Utilizes the sentence-transformer model (\textit{all-MiniLM-L6-v2}\footnote{\url{https://huggingface.co/sentence-transformers/all-MiniLM-L6-v2}}) pre-trained on over 1B English sentence pairs \cite{reimers2019sentence}. 

\item Fine-tuned SBERT: An improved version of the Standard SBERT, incorporating multiple negatives ranking loss with in-batch negatives. The model's training hyperparameters inspired by \citet{decorte2023extreme} are found in Appendix \ref{appendix:training-details}.

\item FinBERT: A BERT model variant specialized for financial texts \cite{liu2021finbert}. FinBERT is not utilized to generate sentence embeddings but rather to understand the financial context at the token level. A mean pooling strategy is used to derive sentence representations.

\item TopoLedgerBERT: Our proposed model. It employs a cosine similarity loss function that acknowledges hierarchical distances between different ledger accounts. Training hyperparameters can be found in the Appendix \ref{appendix:training-details}.
\end{itemize}

\subsection{Performance Metrics}
\label{exp:p-metrics}

In assessing the effectiveness of the proposed TopoLedgerBERT model, we employ four performance metrics: accuracy (Acc), mean reciprocal rank (MRR), and two novel measures, the Mean Misprediction Distance (MMD) and the Mean Overall Distance (MOD). 

The MMD quantifies the average hierarchical distance between the predicted and the actual label for mispredicted instances. A lower MMD implies that, although the prediction is incorrect, the predicted label is still relatively close to the true label in terms of the hierarchical structure of the COA. This metric provides a comprehensive overview of the model's performance in preserving the hierarchical relationships in the COA.

Let $MD(v_i, v_j)$ be the distance of misprediction between the predicted vertex $v_i$ and the true vertex $v_j$, which is defined as the length of the shortest path between $v_i$ and $v_j$ in the COA tree $T$. If the prediction is correct, i.e., $v_i = v_j$, then $MD(v_i, v_j) = 0$. Otherwise, $MD(v_i, v_j) > 0$.

The MMD is then defined as the average misprediction distance over all mispredicted instances in the test set, calculated as follows:

\begin{equation}
MMD = \frac{\sum_{v_i \neq v_j} MD(v_i, v_j)}{|\{(v_i, v_j) | v_i \neq v_j\}|}
\end{equation}

where $\{(v_i, v_j) | v_i \neq v_j\}$ is the set of mispredicted instances.

Similarly, the Mean Overall Distance (MOD) can be defined, which also takes into account correctly predicted instances:
\begin{equation}
MOD = \frac{\sum_{i} MD(v_i, v_j)}{|\{(v_i, v_j)\}|},
\end{equation}
where $\{(v_i, v_j)\}$ is the set of all instances.

\subsection{Experiment 1: Performance Evaluation of TopoLedgerBERT and Impact of Negative Sampling}

In this experiment, we aim to investigate two main research questions: 
\begin{enumerate}[nosep]
    \item How does the TopoLedgerBERT model perform in comparison to other established methods?
    \item Does increasing the number of negative samples in the augmented dataset $D_{aug}$ impact the performance of the TopoLedgerBERT model?
\end{enumerate}
For the second question, we experiment with different values of $K$, where $K$ represents the number of extra negative samples generated per positive sample in $D_{+}$. The values we consider for $K$ are $\{5,10,15,20\}$. We denote the different TopoLedgerBERT models trained on augmented datasets with different negative sample sizes by TopoLedgerBERT@K. 

\begin{table}[htbp]
\centering
\begin{tabular}{llll}
\toprule
\textbf{Metric} && Acc & MRR \\
\midrule
FinBERT && 28.91 & 34.57 \\
Standard SBERT && 44.79 & 53.07 \\
Fine-tuned SBERT && 64.15 & 72.69 \\
TopoLedgerBERT@5 && 65.30 & 72.62 \\
TopoLedgerBERT@10 && 66.13 & 73.23 \\
TopoLedgerBERT@15 && 66.80 & 73.73 \\
TopoLedgerBERT@20 && \textbf{67.01} & \textbf{73.78} \\
\bottomrule
\end{tabular}
\caption{Comparison of the performance of TopoLedgerBERT and benchmark models. Models are evaluated based on Acc and MRR. The best performing model for each metric is in \textbf{bold}.}
\label{table:experiment-1}
\end{table}

The results of the empirical evaluation, summarized in Table \ref{table:experiment-1}, validate the superior performance of our proposed TopoLedgerBERT model over the benchmark methods in terms of accuracy and mean reciprocal rank. 

The FinBERT model, despite being trained on financial data, does not perform as well as the other models. This is likely due to its token-level focus, which does not lend itself to generating meaningful sentence representations. On the other hand, the Standard SBERT and Fine-tuned SBERT models, which are sentence-transformer models, perform significantly better. This highlights the importance of sentence representation methods in the context of ledger account mapping.

When comparing the TopoLedgerBERT@K models with different values of $K$, we observe a general trend of increasing performance with increasing $K$. The best performance is seen at $K=20$, where the model achieves an accuracy of 67.01\% and a MRR score of 73.78\%. This suggests that augmenting the training data with more diverse negative samples containing information about the ledger account taxonomy can effectively enhance the model performance. Learning from negative samples helps the model better distinguish between different ledger accounts and enhance its mapping abilities.

However, the impact of a further increase in $K$ on the model's performance is not straightforward. Excessive negative samples could potentially dilute the impact of the positive samples. Further investigations could be done by testing a wider range of $K$ values.

\subsection{Experiment 2: Hierarchical Relationship Understanding of TopoLedgerBERT}

In this experiment, we evaluate the ability of the TopoLedgerBERT model to capture hierarchical relationships between different ledger accounts. We compare it with the Fine-tuned SBERT model to determine its effectiveness. The comparison is based on two metrics: MOD and MMD.

\begin{table}[htbp]
\centering
\begin{tabular}{lll}
\toprule 
\textbf{Metric} & MOD & MMD \\
\midrule
Fine-tuned SBERT & 1.44 & 4.41 \\
TopoLedgerBERT@20 & \textbf{1.02} & \textbf{3.90} \\
\bottomrule
\end{tabular}
\caption{Comparison of MOD and MMD for Fine-tuned SBERT and TopoLedgerBERT@20. The best results are \textbf{bold}.}
\label{table:experiment-2}
\end{table}

As shown in Table \ref{table:experiment-2}, the TopoLedgerBERT@20 model outperforms the Fine-tuned SBERT model in terms of both MOD and MMD. This suggests that the TopoLedgerBERT model is more effective at capturing hierarchical relationships in the COA and accurately mapping ledger accounts, thus leading to semantically better predictions.

The MOD value for the TopoLedgerBERT@20 model is significantly lower than the value for the Fine-tuned SBERT model. This indicates that the TopoLedgerBERT@20 model's predictions are generally closer to the true labels in the COA, including both correct and incorrect predictions. 

Regarding the MMD value, the TopoLedgerBERT@20 model achieves a lower value. This suggests a more effective use of hierarchical information by the TopoLedgerBERT@20 model, which leads to more accurate predictions, even for mispredicted samples.

Subsequently, we analyze the performance of TopoLedgerBERT@20 by comparing its misprediction distance distribution with that of Fine-tuned SBERT. This analysis provides additional insights into the model's ability to capture and leverage the hierarchical information from the COA. 

\begin{figure}[htbp]
\includegraphics[width=\columnwidth]{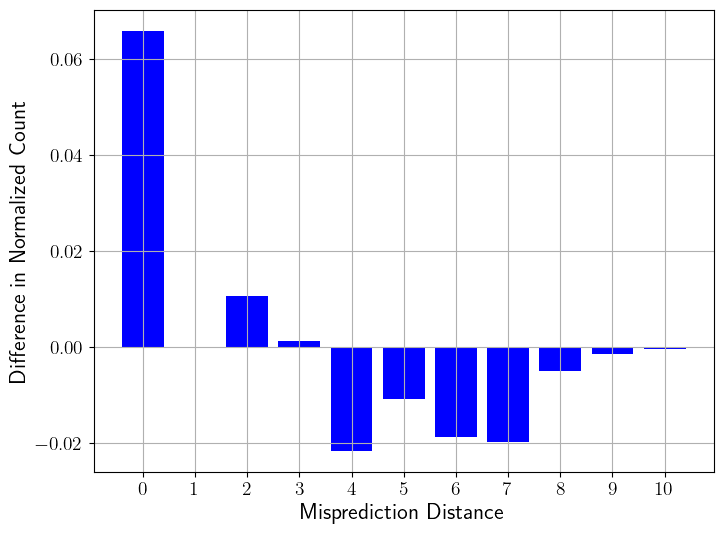}
\caption{Misprediction Distance (MD) difference distribution between TopoLedgerBERT@20 model and Fine-tuned SBERT model.}
\label{fig4}
\end{figure}

The plot, as shown in Figure \ref{fig4}, shows the difference in the distribution of the MD between the TopoLedgerBERT@20 and the Fine-tuned SBERT models. A positive difference indicates that the TopoLedgerBERT@20 model has more predictions at that specific distance, while a negative difference suggests the opposite. Looking at the misprediction distance distribution difference plot, it is clear that the TopoLedgerBERT@20 model outperforms the Fine-tuned SBERT model in terms of both the count of correct predictions and the distribution of mispredictions. 

TopoLedgerBERT@20 outperforms Fine-tuned SBERT by around 6\% in correct predictions. While both models show a similar pattern in one-edge-away predictions, reflecting a preference for specificity, TopoLedgerBERT@20 has more two-edges-away mispredictions. This implies that TopoLedgerBERT@20 leans towards making close-by predictions.

TopoLedgerBERT@20 significantly performs better than Fine-tuned SBERT for predictions that are four edges away or more. This implies that the TopoLedgerBERT@20 model not only increases the count of correct predictions but also reduces the number of far-off predictions, hence preserving the hierarchical relationships in the COA better.

\subsection{Discussion}

The empirical evaluation shows that TopoLedgerBERT outperforms the benchmark models in terms of accuracy and mean reciprocal rank, validating its effectiveness in mapping custom ledger account descriptions to a standardized taxonomy. The increased performance of the TopoLedgerBERT model is attributed to its innovative approach of incorporating hierarchical information from the COA into the training process and employing data augmentation strategies.

Additionally, the TopoLedgerBERT model performs better in capturing hierarchical relationships between different ledger accounts. The lower MOD and MMD scores indicate that the model can make more accurate predictions that are closer to the true labels in the COA.

The results confirm that the TopoLedgerBERT model shows a significant advance toward the ledger account mapping problem. Future research could explore further enhancements to the model, such as incorporating additional contextual information or exploring more complex data augmentation strategies.

\section{Conclusion}
\label{sec:conclusion}

In this paper, we address the significant problem of mapping custom ledger accounts to a standardized COA, a task that is crucial to improving the comparability of financial data between companies. We propose a unique method, TopoLedgerBERT, that leverages hierarchical information in the COA and fine-tuned sentence embeddings to create a mapping solution. Our empirical evaluation shows that TopoLedgerBERT not only improves the mapping accuracy, but also effectively captures the hierarchical relationships between different ledger accounts. Furthermore, the TopoLedgerBERT model shows its effectiveness in addressing the cold-start problem, generally seen when a system must predict for ledger accounts absent from the training dataset. Overcoming this issue not only solves the "multiple charts of accounts problem", as the model is not bound by a predefined set of ledger accounts or a COA, but it also paves the way for more streamlined and efficient automation of COA standardization. The empirical evaluation provides valuable insights into the proposed model's potential benefits and areas for further improvement, ultimately contributing to improving the efficiency and precision of the accounting domain.

\section*{Acknowledgments}

This research received funding from the Flemish government, through Flanders Innovation \& Entrepreneurship (VLAIO, project HBC.2020.2883) and from the Flemish government under the program “Onderzoeksprogramma Artificiële Intelligentie (AI) Vlaanderen”. We also acknowledge the invaluable feedback provided by Simon De Ridder, Karen Dedecker, and Pieter De Koninck of the AI Team at Silverfin.

\bibliography{custom}

\appendix

\section{Details of the Training Process}\label{appendix:training-details}
The strategies for learning sentence representations were executed using the widely used S-BERT\footnote{\url{https://www.sbert.net/docs/package_reference/losses.html}} implementation \cite{reimers2019sentence}. In the multiple negative ranking loss, the 'scale' hyperparameter was kept at the default value of 20. The training was always done for 1 epoch. Positive pairs were mixed randomly into groups of 64. The AdamW optimizer was used as the default optimizer, with a learning rate of 2e-5, and a 'WarmupLinear' learning rate schedule that includes a warm-up period covering 5\% of the training data. Automatic mixed precision was employed to make the training process faster. 

\end{document}